\documentclass[twocolumn,amsmath,prl]{revtex4-1}
\usepackage{amsthm,hyperref,cleveref,braket,amssymb}
\newcommand{\abs}[1]{\left\lvert{#1}\right\rvert}
\newcommand{\prj}[1]{\ket{#1}\bra{#1}}
\newtheorem{thm}{Theorem}
\DeclareMathOperator{\re}{Re}
\DeclareMathOperator{\erf}{erf}
\begin{document}

\title{Anomalous Weak Values Are Proofs of Contextuality}
\author{Matthew F. Pusey}
\email{m@physics.org}
\affiliation{Perimeter Institute for Theoretical Physics, 31 Caroline Street North, Waterloo, ON N2L 2Y5, Canada}
\date{November 12, 2014}
\begin{abstract} 
The average result of a weak measurement of some observable $A$ can, under post-selection of the measured quantum system, exceed the largest eigenvalue of $A$. The nature of weak measurements, as well as the presence of post-selection and hence possible contribution of measurement-disturbance, has led to a long-running debate about whether or not this is surprising. Here, it is shown that such ``anomalous weak values'' are non-classical in a precise sense: a sufficiently weak measurement of one constitutes a proof of contextuality.
This clarifies, for example, which features must be present (and in an experiment, verified) to demonstrate an effect with no satisfying classical explanation.
\end{abstract}
\maketitle

In 1988 Aharonov, Albert and Vaidman explained ``How the result of a measurement of a component of the spin of a spin-$\frac12$ particle can turn out to be 100.'' \cite{weak} Defining the \emph{weak value} of an observable $A$ for a quantum system prepared in state $\ket{\psi}$ and post-selected on giving the first outcome of $\{\prj{\phi}, I - \prj{\phi}\}$,
\begin{equation}
  A_w = \frac{\braket{\phi|A|\psi}}{\braket{\phi|\psi}}, \label{weakdef}
\end{equation}
they exhibited a $\ket{\psi}$ and $\ket{\phi}$ on a qubit for which $Z_w = 100$. The motivation for weak values starts by considering a von Neumann model \cite{von} of the measurement of $A$. The strength of the interaction between the system and ``pointer'' is then drastically reduced, such that the pointer reading is correlated only slightly with $A$. The weak value then arises as an approximation of the average pointer reading to first order in the interaction strength.

Weak values outside the eigenvalue range of $A$ are termed \emph{anomalous}. Aside from possible practical applications (see \cite{review} and references therein), it has been suggested that such values have foundational significance. For example, both their theoretical prediction and experimental observation are said to shed light on ``quantum paradoxes'' \cite{completedesc,threeboxexp,hardyparadox,hardyparadoxexp,cheshire,cheshireexp} and even the nature of time \cite{phystoday}. 

However, there is still no consensus on the most basic question about anomalous weak values: to what extent do they represent a genuinely non-classical effect? The lesser the extent, the more  severe the limitations on their practical and foundational significance.

The arguments that anomalous weak values are non-classical have often been somewhat heuristic, appearing to depend on issues such as the extent to which weak measurements should be called measurements at all \cite{leggett,reply2leggett}. Perhaps the most rigorous evidence provided so far is a connection between anomalous weak values and the failure of a notion of classicality called ``macroscopic realism'' \cite{macrorealism1,macrorealism2,macrorealism3}. On the other hand, classical models have been given that reproduce various aspects of the phenomena \cite{classical1,classical2,classical3}.

The question can be made precise by asking if anomalous weak values constitute proofs of the incompatibility of quantum theory with non-contextual ontological models \cite{cntx}, or equivalently \cite{negativity} if anomalous weak values require negativity in all quasi-probability representations. This was conjectured to be the case in \cite{tollaksen}. Here I will prove it. Interestingly, the proof hinges on two issues already identified in the literature: what do weak measurements measure, and how much do they disturb the system? It transpires that both questions have clear answers in the setting of a non-contextual ontological model, but the particular information-disturbance tradeoff of the weak measurements in quantum theory makes these answers irreconcilable with the anomaly.


Let us begin by specifying exactly what is meant by an anomalous weak value. Inspection of \cref{weakdef} shows that $A_w$ need not be real even though $A$ is Hermitian. A complex number will certainly not be a convex combination of the eigenvalues of $A$, and so this might be seen as surprising. However, the imaginary part of $A_w$ is manifested very differently from the real part \cite{jozsa}. Indeed complex weak values are easily obtained even in the Gaussian subset of quantum mechanics, which has weak measurements (with the same information-tradeoff disturbance utilised here) and yet admits a very natural non-contextual model \cite{erl}. Hence I will call a weak value $A_w$ anomalous only when $\re(A_w)$ is smaller than the smallest eigenvalue of $A$, or larger than the largest eigenvalue of $A$. 

A simplification can be obtained by substituting the spectral decomposition $A = \sum_a a \Pi^{(a)}$ into the RHS of \cref{weakdef} and taking the real part:
\[
  \re(A_w) = \sum_a a \re\left(\frac{\braket{\phi|\Pi^{(a)}|\psi}}{\braket{\phi|\psi}}\right) = \sum_a a \re(\Pi^{(a)}_w).
\]
If we had $0 \leq \re(\Pi^{(a)}_w) \leq 1$ for all $a$ then $A_w$ could not be anomalous. Hence an anomalous weak value for any observable always implies an anomalous weak value for a projector. Since $\sum_a \Pi^{(a)}_w = I_w = 1$, if one projector has $\re(\Pi^{(a)}_w) > 1$ then another must have $\re(\Pi^{(a')}_w) < 0$. In conclusion, without loss of generality we can always take the anomalous weak value to be associated with projector $\Pi$ having $\re(\Pi_w) < 0$.

I will now briefly review the relevant notion of non-contextuality, following \cite{cntx} (where the definitions are motivated and compared to the traditional definition of non-contextuality due to Kochen and Specker \cite{ks}). Assumptions of non-contextuality are constraints on an \emph{ontological model}. I will only need two notions: \emph{measurement non-contextuality}, and \emph{outcome determinism for sharp measurements}. (The latter can be shown to itself follow from the assumption of \emph{preparation non-contextuality} together with some simple facts about quantum theory, see \cite{cntx,robrant} for details.)

Suppose we prepare a quantum system in some way, represented in quantum theory by a state $\ket{\psi}$. In an ontological model the preparation is represented by a probability distribution $p(\lambda)$ over a set of \emph{ontic states} $\Lambda$. Suppose we now implement the POVM $\{E_k\}$. In a measurement non-contextual model, this is represented by a conditional probability distribution $\{ p(E_k|\lambda) \}$. The assumption of measurement non-contextuality is what allows us to write $p(E_k|\lambda)$ as a function of the effect $E_k$ and the ontic state $\lambda$ only, with no dependence on other things (``contexts''), such as the other elements of the POVM or details of how the POVM was implemented. Outcome determinism for sharp measurements is the assumption that $p(\Pi|\lambda) \in \{0,1\}$ for all projectors $\Pi$ and ontic states $\lambda$, so that any inability to predict the outcome of a projective measurement is due purely to ignorance of $\lambda$.

The final requirement, for any ontological model, is that when we marginalise over the ontic states, the model must reproduce the predictions of quantum theory:
\begin{equation}
  \braket{\psi|E_k|\psi} = \int_\Lambda p(E_k|\lambda)p(\lambda) d\lambda. \label{reproduce}
\end{equation}

We can now state the main result, identifying certain features in the measurement of anomalous weak values that, taken together, defy non-contextual explanation.
\begin{thm}\label{mainthm}
  Suppose we have states $\ket{\phi}$ and $\ket{\psi}$, and a generalized measurement \cite{nc} $\{M_x\}_{x \in \mathbb{R}}$, such that
  \begin{enumerate}
    \item The pre- and post-selection are non-orthogonal, i.e.
       \begin{equation}
       p_\phi := \abs{\braket{\phi|\psi}}^2 > 0,
       \end{equation}
    \item The POVM is a projector plus unbiased noise, i.e.
      \begin{equation}
	E_x := M_x^\dagger M_x = p_n(x-1)\Pi + p_n(x)\tilde\Pi\label{noiseprop}
      \end{equation}
      for some projector $\Pi$, $\tilde\Pi = I - \Pi$, and probability distribution $p_n(x)$ with median $x = 0$,
    \item We can define a probability $p_d$ (the ``probability of disturbance'') such that
      \begin{equation}
	S := \int_{-\infty}^\infty M_x^\dagger \prj{\phi} M_x dx = (1-p_d)\prj{\phi} + p_d E_d\label{disturbprop}
      \end{equation}
      for some POVM $\{E_d, I-E_d\}$, and
    \item The values of $x$ under the pre- and post-selection have a negative bias that ``outweighs'' $p_d$, i.e. \footnote{Notice that although $p_-$ is a combination of operationally defined quantities, it is not exactly the probability of getting a negative $x$ under the pre- and post-selection. To obtain this, instead of dividing by $p_\phi$ one would have to divide by $\Braket{\psi|S|\psi}= (1-p_d)p_\phi + p_d\Braket{\psi|E_d|\psi}$, making the analysis slightly more complicated (but still tractable).}
      \begin{equation}
	p_- := \frac{1}{p_\phi}\int_{-\infty}^0 \abs{\Braket{\phi|M_x|\psi}}^2 dx > \frac12 + \frac{p_d}{p_\phi}.\label{biasprop}
      \end{equation}
   \end{enumerate}

   Then there is no measurement non-contextual ontological model for the preparation of $\ket{\psi}$, measurement of $\{M_x\}$, and post-selection of $\ket{\phi}$ satisfying outcome determinism for sharp measurements.
\end{thm}
(Showing that operators $\{M_x\}$ with these properties actually exist whenever we have a $\ket{\psi}$, $\ket{\phi}$ and $\Pi$ with $\re(\Pi_w) < 0$ is a routine calculation in the theory of weak measurement \cite{weak,jozsa,wiseweak}, postponed until later. Loosely speaking, if $g \ll 1$ is the strength of the measurement then to leading order $(p_- - \frac12) \sim g$ whereas $p_d \sim g^2$.)

\begin{proof}
Suppose such an ontological model exists. We can consider the weak measurement $\{M_x\}$ followed by the projective measurement $\{\prj{\phi}, I-\prj{\phi}\}$ as one ``consolidated measurement'', represented by the POVM $\{S_x \} \cup \{ F_x\}$, where $S_x = M_x^\dagger \prj{\phi} M_x$ and $F_x = M_x^\dagger (I - \prj{\phi}) M_x$.  The key question is how the $\{S_x\}$ are represented in the model, because \cref{reproduce} gives
\begin{equation}
  \abs{\braket{\phi|M_x|\psi}}^2 = \Braket{\psi|S_x|\psi} = \int_\Lambda p(S_x|\lambda)p(\lambda)d\lambda. \label{ppsreproduce}
\end{equation}

Let us consider two methods for implementing the POVM $\{E_x\}$. By the assumption of measurement non-contextuality they must both lead to the same $p(E_x|\lambda)$. The first method is to implement the consolidated measurement and then ignore the result of the post-selection, giving $p(E_x | \lambda) = p(S_x|\lambda) + p(F_x|\lambda)$. The second method, according to \cref{noiseprop}, is to measure $\{\Pi, \tilde\Pi\}$ and then classically sample from $p_n(x-1)$ or $p_n(x)$ as appropriate. Hence we also have $p(E_x | \lambda) = p_n(x-1)p(\Pi|\lambda) + p_n(x)p(\tilde\Pi|\lambda)$. Since the median of $p_n(x)$ is $0$ we have $\int_{-\infty}^0 p_n(x-1)dx \leq \int_{-\infty}^0 p_n(x)dx = \frac12$. Combining this with $p(S_x|\lambda) \leq p(E_x|\lambda)$ from the first method, we have
\begin{equation}
  \int_{-\infty}^0 p(S_x|\lambda) dx \leq \int_{-\infty}^0 p(E_x|\lambda) dx \leq \frac12.\label{ontnobias}
\end{equation}


Next, we apply the assumption of measurement non-contextuality to the POVM $\{ S, I - S \}$. One way to implement this is to use the consolidated measurement and ignore $x$, hence $p(S|\lambda) = \int_{-\infty}^\infty p(S_x|\lambda)dx$. A second way, according to \cref{disturbprop}, is to measure $\{\prj{\phi}, I - \prj{\phi}\}$ with probability $1-p_d$ and $\{E_d, I-E_d\}$ with probability $p_d$. Hence $p(S|\lambda) = (1-p_d)p(\prj{\phi}|\lambda) + p_d p(E_d|\lambda)$.

Finally, we calculate the model's prediction for $p_-$. Using outcome determinism for the sharp measurement $\{\prj{\phi}, I-\prj{\phi}\}$ we can partition $\Lambda$ into $\{\Lambda_0,\Lambda_1\}$ where $p(\prj{\phi}|\lambda)=i$ for $\lambda \in \Lambda_i$. From the above we have that $\int_{-\infty}^0 p(S_x|\lambda)dx \leq p(S|\lambda) \leq p_d$ on $\Lambda_0$. Hence splitting the RHS of \eqref{ppsreproduce} into integrals over $\Lambda_0$ and $\Lambda_1$ and integrating over $x<0$ gives
\[
  \int_{-\infty}^0 \abs{\braket{\phi|M_x|\psi}}^2dx \leq \int_{-\infty}^0\int_{\Lambda_1}p(S_x|\lambda)p(\lambda)d\lambda dx + p_d.
  \]
Applying \cref{ontnobias} and recalling that \eqref{reproduce} gives $\int_{\Lambda_1}p(\lambda)d\lambda = \int_\Lambda p(\prj{\phi}|\lambda)p(\lambda)d\lambda = \abs{\braket{\phi|\psi}}^2 = p_\phi$ this gives
\begin{equation}
  \frac{1}{p_\phi}\int_{-\infty}^0 \abs{\braket{\phi|M_x|\psi}}^2dx \leq \frac12  + \frac{p_d}{p_\phi}.\label{numbound}
\end{equation}
in contradiction to \cref{biasprop}.
\end{proof}

As promised, I will now confirm that a projector $\Pi$ with $\re(\Pi_w) < 0$ implies the existence of a measurement $\{M_x\}$ with the properties assumed in \cref{mainthm}.

Similarly to \cite{weak}, the measurement begins by preparing a probe system in the Gaussian state $\ket{\Psi} = N \int_{-\infty}^\infty \exp(-x^2/2\sigma^2)\ket{x}dx$, with $N=(\pi\sigma^2)^{-1/4}$. This interacts with the system via the unitary (with $\hbar = 1$)
\begin{equation}
U = \exp(-i\Pi P) = \exp(-iP)\Pi + \tilde\Pi,
\end{equation}
which defines our units of momentum and hence length, and then the probe is projectively measured in the $\{\prj{x}\}$ basis. On the system this is a generalised measurement with $M_x = \braket{x|U|\Psi}$. Recalling that $P$ generates translations we have
\begin{equation}
  M_x = N\exp\left(-\frac{(x-1)^2}{2\sigma^2}\right)\Pi + N\exp\left(-\frac{x^2}{2\sigma^2}\right)\tilde\Pi.
\end{equation}

This becomes a projective measurement in the limit $\sigma \to 0$, whereas it is known as a weak measurement for large $\sigma$. We can now calculate
\begin{equation}
  E_x = M_x^\dagger M_x = p_n(x-1)\Pi + p_n(x)\tilde\Pi
\end{equation}
where $p_n(x) = N^2\exp(-x^2/\sigma^2)$ has median $x=0$. Recalling that $p_n(x)$ is normalised and defining
\begin{equation}
  \Delta := \int_{-\infty}^\infty\sqrt{p_n(x-1)p_n(x)}dx = \exp\left(-\frac{1}{4\sigma^2}\right),
\end{equation}
we obtain
\begin{multline}
  S = \int_{-\infty}^\infty M_x^\dagger \prj{\phi} M_x dx \\= \Pi\prj{\phi}\Pi + \tilde\Pi\prj{\phi}\tilde\Pi + \Delta (\Pi\prj{\phi}\tilde\Pi + \tilde\Pi\prj{\phi}\Pi)
  \\= \frac{1+\Delta}2\prj{\phi} + \frac{1-\Delta}2(\Pi-\tilde\Pi)\prj{\phi}(\Pi-\tilde\Pi).
\end{multline}
Setting $p_d = \frac{1-\Delta}2$ and $E_d = (\Pi-\tilde\Pi)\prj{\phi}(\Pi-\tilde\Pi)$ (which is a projector) we have \cref{disturbprop}.

Finally we need to calculate
\begin{multline}
  p_- = \frac{1}{p_\phi}\int_{-\infty}^0\abs{\braket{\phi|M_x|\psi}}^2dx \\= A\abs{\Pi_w}^2 + B\lvert\tilde\Pi_w\rvert^2 + 2C\re(\Pi_w\tilde\Pi_w^*),
\end{multline}
where we have recalled \cref{weakdef} and defined the integrals
\begin{align}
  A &= \int_{-\infty}^0p_n(x-1)dx = \frac12\left(1 - \erf\left(\frac{1}{\sigma}\right)\right),\\
  B &= \int_{-\infty}^0p_n(x)dx = \frac12,\\
  C &= \int_{-\infty}^0\sqrt{p_n(x-1)p_n(x)}dx \nonumber\\&= \frac12\exp\left(-\frac{1}{4\sigma^2}\right)\left(1-\erf\left( \frac{1}{2\sigma} \right)\right).
\end{align}

Expanding around $1/\sigma = 0$ we find $A \approx \frac12 -\frac{1}{\sqrt\pi\sigma}$ and $C \approx \frac12 - \frac{1}{2\sqrt{\pi}\sigma}$. Since $\Pi_w + \tilde\Pi_w = I_w = 1$ this gives
\begin{equation}
  p_- \approx \frac12 - \frac{1}{\sqrt{\pi}\sigma}\re(\Pi_w).
\end{equation}
Meanwhile to leading order $p_d \approx \frac{1}{8\sigma^2}$. Hence, provided $\re(\Pi_w) < 0$, for sufficiently large $\sigma$ we will satisfy \cref{biasprop}. It is worth emphasising that no approximations were made in the proof of \cref{mainthm}, and in a concrete case one can simply plug values of $\sigma$ into the exact formulas above to verify \cref{biasprop}.


I will conclude by outlining three interconnected lessons from \cref{mainthm}. The first is a classification of how anomalous weak values could arise in an ontological model. One possibility (perhaps the most common realist interpretation of anomalous weak values) is that some ontic states are pre-disposed to manifest such values, in violation of the first application of measurement non-contextuality using \cref{noiseprop}. Alternatively (along the lines of \cite{classical3}) the weak measurement may disturb the system much more than the quantum formalism would suggest, in violation of the second application. The final possibility is that the post-selection is not represented outcome deterministically (as in the interpretation where the ontic state is simply the quantum state) and so fails to identify a particular set of ontic states.

The second lesson is that a large number of aspects of the manifestation of anomalous weak values seem to be involved in preventing non-contextual explanation. The ``anomaly'' itself is only one ingredient. Some others may have been anticipated, such as the favourable information-disturbance tradeoff of weak measurements. But some seem somewhat surprising, for example the importance of the post-selection being a projective measurement.

The final lesson is an indication of what it would take for an experiment involving anomalous weak values to exclude non-contextual theories that would provide a good classical explanation. Merely observing ``anomalous pointer readings'' under pre- and post-selection is far from sufficient. Most fundamentally, the experiment must show that the probabilities in the statement of \cref{mainthm} really are the probabilities of discrete events, rather than mere (normalised) intensities. An experiment consistent with a classical field theory, so far the most common way to observe weak values, is therefore not sufficient \footnote{This is because the analysis presented here, like any proof of contextuality, is for an ontological model that produces individual measurement results with the correct probabilities. This requirement immediately rules out a straightforward field ontology that, whilst perhaps offering interesting explanations of the weak value \cite{classical2,field1,field2}, only produces intensities. To exclude non-contextual explanation, an experiment based on fields would have to justify this requirement by working at the level of single field quanta. Compare with the classic ``double-slit experiment'': whilst an interference pattern in intensities has a simple classical explanation in terms of fields, the same interference pattern in what are unambiguously probabilities defies classical intuitions}. One would also need to provide evidence for an operational version of \cref{noiseprop,disturbprop}. Notice that these would be statements about how the weak measurement works on \emph{all} preparations, not just the one corresponding to $\ket{\psi}$. Furthermore, one would need an operational counterpart to the inference from preparation non-contextuality to outcome determinism for the post-selection measurement, perhaps by implementing preparations that make the post-selection highly predictable (see \cite{ravi} for how this can be done in more traditional proofs of contextuality). Turning these ideas into a concrete experimental proposal is an interesting avenue for future work.

\begin{acknowledgments}
Thanks to Aharon Brodutch, Joshua Combes, Chris Ferrie, Ravi Kunjwal and Matt Leifer for useful discussions. I am particularly indebted to Matt for help in analysing measurement-disturbance, and to Aharon for bringing the issue of intensities versus probabilities to my attention.  Research at Perimeter Institute is supported in part by the Government of Canada through NSERC and by the Province of Ontario through MRI.
\end{acknowledgments}

\bibliography{pps}

\begin{thebibliography}{32}%
\makeatletter
\providecommand \@ifxundefined [1]{%
 \@ifx{#1\undefined}
}%
\providecommand \@ifnum [1]{%
 \ifnum #1\expandafter \@firstoftwo
 \else \expandafter \@secondoftwo
 \fi
}%
\providecommand \@ifx [1]{%
 \ifx #1\expandafter \@firstoftwo
 \else \expandafter \@secondoftwo
 \fi
}%
\providecommand \natexlab [1]{#1}%
\providecommand \enquote  [1]{``#1''}%
\providecommand \bibnamefont  [1]{#1}%
\providecommand \bibfnamefont [1]{#1}%
\providecommand \citenamefont [1]{#1}%
\providecommand \href@noop [0]{\@secondoftwo}%
\providecommand \href [0]{\begingroup \@sanitize@url \@href}%
\providecommand \@href[1]{\@@startlink{#1}\@@href}%
\providecommand \@@href[1]{\endgroup#1\@@endlink}%
\providecommand \@sanitize@url [0]{\catcode `\\12\catcode `\$12\catcode
  `\&12\catcode `\#12\catcode `\^12\catcode `\_12\catcode `\%12\relax}%
\providecommand \@@startlink[1]{}%
\providecommand \@@endlink[0]{}%
\providecommand \url  [0]{\begingroup\@sanitize@url \@url }%
\providecommand \@url [1]{\endgroup\@href {#1}{\urlprefix }}%
\providecommand \urlprefix  [0]{URL }%
\providecommand \Eprint [0]{\href }%
\providecommand \doibase [0]{http://dx.doi.org/}%
\providecommand \selectlanguage [0]{\@gobble}%
\providecommand \bibinfo  [0]{\@secondoftwo}%
\providecommand \bibfield  [0]{\@secondoftwo}%
\providecommand \translation [1]{[#1]}%
\providecommand \BibitemOpen [0]{}%
\providecommand \bibitemStop [0]{}%
\providecommand \bibitemNoStop [0]{.\EOS\space}%
\providecommand \EOS [0]{\spacefactor3000\relax}%
\providecommand \BibitemShut  [1]{\csname bibitem#1\endcsname}%
\let\auto@bib@innerbib\@empty
\bibitem [{\citenamefont {Aharonov}\ \emph {et~al.}(1988)\citenamefont
  {Aharonov}, \citenamefont {Albert},\ and\ \citenamefont {Vaidman}}]{weak}%
  \BibitemOpen
  \bibfield  {author} {\bibinfo {author} {\bibfnamefont {Y.}~\bibnamefont
  {Aharonov}}, \bibinfo {author} {\bibfnamefont {D.~Z.}\ \bibnamefont
  {Albert}}, \ and\ \bibinfo {author} {\bibfnamefont {L.}~\bibnamefont
  {Vaidman}},\ }\href {\doibase 10.1103/PhysRevLett.60.1351} {\bibfield
  {journal} {\bibinfo  {journal} {Phys. Rev. Lett.}\ }\textbf {\bibinfo
  {volume} {60}},\ \bibinfo {pages} {1351} (\bibinfo {year} {1988})},\ \bibinfo
  {note} {\url{http://www.tau.ac.il/~vaidman/lvhp/m8.pdf}}\BibitemShut
  {NoStop}%
\bibitem [{\citenamefont {von Neumann}(1996)}]{von}%
  \BibitemOpen
  \bibfield  {author} {\bibinfo {author} {\bibfnamefont {J.}~\bibnamefont {von
  Neumann}},\ }\href@noop {} {\emph {\bibinfo {title} {Mathematical Foundations
  of Quantum Mechanics}}}\ (\bibinfo  {publisher} {Princeton University
  Press},\ \bibinfo {year} {1996})\BibitemShut {NoStop}%
\bibitem [{\citenamefont {Dressel}\ \emph
  {et~al.}(2014{\natexlab{a}})\citenamefont {Dressel}, \citenamefont {Malik},
  \citenamefont {Miatto}, \citenamefont {Jordan},\ and\ \citenamefont
  {Boyd}}]{review}%
  \BibitemOpen
  \bibfield  {author} {\bibinfo {author} {\bibfnamefont {J.}~\bibnamefont
  {Dressel}}, \bibinfo {author} {\bibfnamefont {M.}~\bibnamefont {Malik}},
  \bibinfo {author} {\bibfnamefont {F.~M.}\ \bibnamefont {Miatto}}, \bibinfo
  {author} {\bibfnamefont {A.~N.}\ \bibnamefont {Jordan}}, \ and\ \bibinfo
  {author} {\bibfnamefont {R.~W.}\ \bibnamefont {Boyd}},\ }\href {\doibase
  10.1103/RevModPhys.86.307} {\bibfield  {journal} {\bibinfo  {journal} {Rev.
  Mod. Phys.}\ }\textbf {\bibinfo {volume} {86}},\ \bibinfo {pages} {307}
  (\bibinfo {year} {2014}{\natexlab{a}})},\ \Eprint
  {http://arxiv.org/abs/1305.7154} {arXiv:1305.7154} \BibitemShut {NoStop}%
\bibitem [{\citenamefont {Aharonov}\ and\ \citenamefont
  {Vaidman}(1991)}]{completedesc}%
  \BibitemOpen
  \bibfield  {author} {\bibinfo {author} {\bibfnamefont {Y.}~\bibnamefont
  {Aharonov}}\ and\ \bibinfo {author} {\bibfnamefont {L.}~\bibnamefont
  {Vaidman}},\ }\href {\doibase 10.1088/0305-4470/24/10/018} {\bibfield
  {journal} {\bibinfo  {journal} {J. Phys. A: Math. Gen.}\ }\textbf {\bibinfo
  {volume} {24}},\ \bibinfo {pages} {2315} (\bibinfo {year}
  {1991})}\BibitemShut {NoStop}%
\bibitem [{\citenamefont {Resch}\ \emph {et~al.}(2004)\citenamefont {Resch},
  \citenamefont {Lundeen},\ and\ \citenamefont {Steinberg}}]{threeboxexp}%
  \BibitemOpen
  \bibfield  {author} {\bibinfo {author} {\bibfnamefont {K.~J.}\ \bibnamefont
  {Resch}}, \bibinfo {author} {\bibfnamefont {J.~S.}\ \bibnamefont {Lundeen}},
  \ and\ \bibinfo {author} {\bibfnamefont {A.~M.}\ \bibnamefont {Steinberg}},\
  }\href {\doibase 10.1016/j.physleta.2004.02.042} {\bibfield  {journal}
  {\bibinfo  {journal} {Phys. Lett. A}\ }\textbf {\bibinfo {volume} {324}},\
  \bibinfo {pages} {125 } (\bibinfo {year} {2004})},\ \Eprint
  {http://arxiv.org/abs/quant-ph/0310091} {arXiv:quant-ph/0310091} \BibitemShut
  {NoStop}%
\bibitem [{\citenamefont {Aharonov}\ \emph {et~al.}(2002)\citenamefont
  {Aharonov}, \citenamefont {Botero}, \citenamefont {Popescu}, \citenamefont
  {Reznik},\ and\ \citenamefont {Tollaksen}}]{hardyparadox}%
  \BibitemOpen
  \bibfield  {author} {\bibinfo {author} {\bibfnamefont {Y.}~\bibnamefont
  {Aharonov}}, \bibinfo {author} {\bibfnamefont {A.}~\bibnamefont {Botero}},
  \bibinfo {author} {\bibfnamefont {S.}~\bibnamefont {Popescu}}, \bibinfo
  {author} {\bibfnamefont {B.}~\bibnamefont {Reznik}}, \ and\ \bibinfo {author}
  {\bibfnamefont {J.}~\bibnamefont {Tollaksen}},\ }\href {\doibase
  10.1016/S0375-9601(02)00986-6} {\bibfield  {journal} {\bibinfo  {journal}
  {Physics Letters A}\ }\textbf {\bibinfo {volume} {301}},\ \bibinfo {pages}
  {130 } (\bibinfo {year} {2002})},\ \Eprint
  {http://arxiv.org/abs/quant-ph/0104062} {arXiv:quant-ph/0104062} \BibitemShut
  {NoStop}%
\bibitem [{\citenamefont {Lundeen}\ and\ \citenamefont
  {Steinberg}(2009)}]{hardyparadoxexp}%
  \BibitemOpen
  \bibfield  {author} {\bibinfo {author} {\bibfnamefont {J.~S.}\ \bibnamefont
  {Lundeen}}\ and\ \bibinfo {author} {\bibfnamefont {A.~M.}\ \bibnamefont
  {Steinberg}},\ }\href {\doibase 10.1103/PhysRevLett.102.020404} {\bibfield
  {journal} {\bibinfo  {journal} {Phys. Rev. Lett.}\ }\textbf {\bibinfo
  {volume} {102}},\ \bibinfo {pages} {020404} (\bibinfo {year} {2009})},\
  \Eprint {http://arxiv.org/abs/0810.4229} {arXiv:0810.4229} \BibitemShut
  {NoStop}%
\bibitem [{\citenamefont {Aharonov}\ \emph {et~al.}(2013)\citenamefont
  {Aharonov}, \citenamefont {Popescu}, \citenamefont {Rohrlich},\ and\
  \citenamefont {Skrzypczyk}}]{cheshire}%
  \BibitemOpen
  \bibfield  {author} {\bibinfo {author} {\bibfnamefont {Y.}~\bibnamefont
  {Aharonov}}, \bibinfo {author} {\bibfnamefont {S.}~\bibnamefont {Popescu}},
  \bibinfo {author} {\bibfnamefont {D.}~\bibnamefont {Rohrlich}}, \ and\
  \bibinfo {author} {\bibfnamefont {P.}~\bibnamefont {Skrzypczyk}},\ }\href
  {\doibase 10.1088/1367-2630/15/11/113015} {\bibfield  {journal} {\bibinfo
  {journal} {New J. Phys.}\ }\textbf {\bibinfo {volume} {15}},\ \bibinfo
  {pages} {113015} (\bibinfo {year} {2013})},\ \Eprint
  {http://arxiv.org/abs/arXiv:1202.0631} {arXiv:arXiv:1202.0631} \BibitemShut
  {NoStop}%
\bibitem [{\citenamefont {Denkmayr}\ \emph {et~al.}(2014)\citenamefont
  {Denkmayr}, \citenamefont {Geppert}, \citenamefont {Sponar}, \citenamefont
  {Lemmel}, \citenamefont {Matzkin}, \citenamefont {Tollaksen},\ and\
  \citenamefont {Hasegawa}}]{cheshireexp}%
  \BibitemOpen
  \bibfield  {author} {\bibinfo {author} {\bibfnamefont {T.}~\bibnamefont
  {Denkmayr}}, \bibinfo {author} {\bibfnamefont {H.}~\bibnamefont {Geppert}},
  \bibinfo {author} {\bibfnamefont {S.}~\bibnamefont {Sponar}}, \bibinfo
  {author} {\bibfnamefont {H.}~\bibnamefont {Lemmel}}, \bibinfo {author}
  {\bibfnamefont {A.}~\bibnamefont {Matzkin}}, \bibinfo {author} {\bibfnamefont
  {J.}~\bibnamefont {Tollaksen}}, \ and\ \bibinfo {author} {\bibfnamefont
  {Y.}~\bibnamefont {Hasegawa}},\ }\href {\doibase 10.1038/ncomms5492}
  {\bibfield  {journal} {\bibinfo  {journal} {Nat. Commun.}\ }\textbf {\bibinfo
  {volume} {5}},\ \bibinfo {pages} {4492} (\bibinfo {year} {2014})},\ \Eprint
  {http://arxiv.org/abs/1312.3775} {arXiv:1312.3775} \BibitemShut {NoStop}%
\bibitem [{\citenamefont {Aharonov}\ \emph {et~al.}(2010)\citenamefont
  {Aharonov}, \citenamefont {Popescu},\ and\ \citenamefont
  {Tollaksen}}]{phystoday}%
  \BibitemOpen
  \bibfield  {author} {\bibinfo {author} {\bibfnamefont {Y.}~\bibnamefont
  {Aharonov}}, \bibinfo {author} {\bibfnamefont {S.}~\bibnamefont {Popescu}}, \
  and\ \bibinfo {author} {\bibfnamefont {J.}~\bibnamefont {Tollaksen}},\ }\href
  {\doibase 10.1063/1.3518209} {\bibfield  {journal} {\bibinfo  {journal}
  {Phys. Today}\ }\textbf {\bibinfo {volume} {63}},\ \bibinfo {pages} {27}
  (\bibinfo {year} {2010})}\BibitemShut {NoStop}%
\bibitem [{\citenamefont {Leggett}(1989)}]{leggett}%
  \BibitemOpen
  \bibfield  {author} {\bibinfo {author} {\bibfnamefont {A.~J.}\ \bibnamefont
  {Leggett}},\ }\href {\doibase 10.1103/PhysRevLett.62.2325} {\bibfield
  {journal} {\bibinfo  {journal} {Phys. Rev. Lett.}\ }\textbf {\bibinfo
  {volume} {62}},\ \bibinfo {pages} {2325} (\bibinfo {year}
  {1989})}\BibitemShut {NoStop}%
\bibitem [{\citenamefont {Aharonov}\ and\ \citenamefont
  {Vaidman}(1989)}]{reply2leggett}%
  \BibitemOpen
  \bibfield  {author} {\bibinfo {author} {\bibfnamefont {Y.}~\bibnamefont
  {Aharonov}}\ and\ \bibinfo {author} {\bibfnamefont {L.}~\bibnamefont
  {Vaidman}},\ }\href {\doibase 10.1103/PhysRevLett.62.2327} {\bibfield
  {journal} {\bibinfo  {journal} {Phys. Rev. Lett.}\ }\textbf {\bibinfo
  {volume} {62}},\ \bibinfo {pages} {2327} (\bibinfo {year} {1989})},\ \bibinfo
  {note} {\url{http://www.tau.ac.il/~vaidman/lvhp/m9.pdf}}\BibitemShut
  {NoStop}%
\bibitem [{\citenamefont {Williams}\ and\ \citenamefont
  {Jordan}(2008)}]{macrorealism1}%
  \BibitemOpen
  \bibfield  {author} {\bibinfo {author} {\bibfnamefont {N.~S.}\ \bibnamefont
  {Williams}}\ and\ \bibinfo {author} {\bibfnamefont {A.~N.}\ \bibnamefont
  {Jordan}},\ }\href {\doibase 10.1103/PhysRevLett.100.026804} {\bibfield
  {journal} {\bibinfo  {journal} {Phys. Rev. Lett.}\ }\textbf {\bibinfo
  {volume} {100}},\ \bibinfo {pages} {026804} (\bibinfo {year} {2008})},\
  \Eprint {http://arxiv.org/abs/0707.3427} {arXiv:0707.3427} \BibitemShut
  {NoStop}%
\bibitem [{\citenamefont {Goggin}\ \emph {et~al.}(2011)\citenamefont {Goggin},
  \citenamefont {Almeida}, \citenamefont {Barbieri}, \citenamefont {Lanyon},
  \citenamefont {O’Brien}, \citenamefont {White},\ and\ \citenamefont
  {Pryde}}]{macrorealism2}%
  \BibitemOpen
  \bibfield  {author} {\bibinfo {author} {\bibfnamefont {M.~E.}\ \bibnamefont
  {Goggin}}, \bibinfo {author} {\bibfnamefont {M.~P.}\ \bibnamefont {Almeida}},
  \bibinfo {author} {\bibfnamefont {M.}~\bibnamefont {Barbieri}}, \bibinfo
  {author} {\bibfnamefont {B.~P.}\ \bibnamefont {Lanyon}}, \bibinfo {author}
  {\bibfnamefont {J.~L.}\ \bibnamefont {O’Brien}}, \bibinfo {author}
  {\bibfnamefont {A.~G.}\ \bibnamefont {White}}, \ and\ \bibinfo {author}
  {\bibfnamefont {G.~J.}\ \bibnamefont {Pryde}},\ }\href {\doibase
  10.1073/pnas.1005774108} {\bibfield  {journal} {\bibinfo  {journal} {Proc.
  Natl. Acad. Sci. USA}\ }\textbf {\bibinfo {volume} {108}},\ \bibinfo {pages}
  {1256} (\bibinfo {year} {2011})},\ \Eprint {http://arxiv.org/abs/0907.1679}
  {arXiv:0907.1679} \BibitemShut {NoStop}%
\bibitem [{\citenamefont {Dressel}\ \emph {et~al.}(2011)\citenamefont
  {Dressel}, \citenamefont {Broadbent}, \citenamefont {Howell},\ and\
  \citenamefont {Jordan}}]{macrorealism3}%
  \BibitemOpen
  \bibfield  {author} {\bibinfo {author} {\bibfnamefont {J.}~\bibnamefont
  {Dressel}}, \bibinfo {author} {\bibfnamefont {C.~J.}\ \bibnamefont
  {Broadbent}}, \bibinfo {author} {\bibfnamefont {J.~C.}\ \bibnamefont
  {Howell}}, \ and\ \bibinfo {author} {\bibfnamefont {A.~N.}\ \bibnamefont
  {Jordan}},\ }\href {\doibase 10.1103/PhysRevLett.106.040402} {\bibfield
  {journal} {\bibinfo  {journal} {Phys. Rev. Lett.}\ }\textbf {\bibinfo
  {volume} {106}},\ \bibinfo {pages} {040402} (\bibinfo {year} {2011})},\
  \Eprint {http://arxiv.org/abs/1101.4917} {arXiv:1101.4917} \BibitemShut
  {NoStop}%
\bibitem [{\citenamefont {Dressel}\ and\ \citenamefont
  {Jordan}(2012)}]{classical1}%
  \BibitemOpen
  \bibfield  {author} {\bibinfo {author} {\bibfnamefont {J.}~\bibnamefont
  {Dressel}}\ and\ \bibinfo {author} {\bibfnamefont {A.~N.}\ \bibnamefont
  {Jordan}},\ }\href {\doibase 10.1103/PhysRevA.85.022123} {\bibfield
  {journal} {\bibinfo  {journal} {Phys. Rev. A}\ }\textbf {\bibinfo {volume}
  {85}},\ \bibinfo {pages} {022123} (\bibinfo {year} {2012})},\ \Eprint
  {http://arxiv.org/abs/1110.0418} {arXiv:1110.0418} \BibitemShut {NoStop}%
\bibitem [{\citenamefont {Bliokh}\ \emph {et~al.}(2013)\citenamefont {Bliokh},
  \citenamefont {Bekshaev}, \citenamefont {Kofman},\ and\ \citenamefont
  {Nori}}]{classical2}%
  \BibitemOpen
  \bibfield  {author} {\bibinfo {author} {\bibfnamefont {K.~Y.}\ \bibnamefont
  {Bliokh}}, \bibinfo {author} {\bibfnamefont {A.~Y.}\ \bibnamefont
  {Bekshaev}}, \bibinfo {author} {\bibfnamefont {A.~G.}\ \bibnamefont
  {Kofman}}, \ and\ \bibinfo {author} {\bibfnamefont {F.}~\bibnamefont
  {Nori}},\ }\href {\doibase 10.1088/1367-2630/15/7/073022} {\bibfield
  {journal} {\bibinfo  {journal} {New J. Phys.}\ }\textbf {\bibinfo {volume}
  {15}},\ \bibinfo {pages} {073022} (\bibinfo {year} {2013})},\ \Eprint
  {http://arxiv.org/abs/1304.1276} {arXiv:1304.1276} \BibitemShut {NoStop}%
\bibitem [{\citenamefont {Ferrie}\ and\ \citenamefont
  {Combes}(2014)}]{classical3}%
  \BibitemOpen
  \bibfield  {author} {\bibinfo {author} {\bibfnamefont {C.}~\bibnamefont
  {Ferrie}}\ and\ \bibinfo {author} {\bibfnamefont {J.}~\bibnamefont
  {Combes}},\ }\href {\doibase 10.1103/PhysRevLett.113.120404} {\bibfield
  {journal} {\bibinfo  {journal} {Phys. Rev. Lett.}\ }\textbf {\bibinfo
  {volume} {113}},\ \bibinfo {pages} {120404} (\bibinfo {year} {2014})},\
  \Eprint {http://arxiv.org/abs/1403.2362} {arXiv:1403.2362} \BibitemShut
  {NoStop}%
\bibitem [{\citenamefont {Spekkens}(2005)}]{cntx}%
  \BibitemOpen
  \bibfield  {author} {\bibinfo {author} {\bibfnamefont {R.~W.}\ \bibnamefont
  {Spekkens}},\ }\href {\doibase 10.1103/PhysRevA.71.052108} {\bibfield
  {journal} {\bibinfo  {journal} {Phys. Rev. A}\ }\textbf {\bibinfo {volume}
  {71}},\ \bibinfo {pages} {052108} (\bibinfo {year} {2005})},\ \Eprint
  {http://arxiv.org/abs/quant-ph/0406166} {arXiv:quant-ph/0406166} \BibitemShut
  {NoStop}%
\bibitem [{\citenamefont {Spekkens}(2008)}]{negativity}%
  \BibitemOpen
  \bibfield  {author} {\bibinfo {author} {\bibfnamefont {R.~W.}\ \bibnamefont
  {Spekkens}},\ }\href {\doibase 10.1103/PhysRevLett.101.020401} {\bibfield
  {journal} {\bibinfo  {journal} {Phys. Rev. Lett.}\ }\textbf {\bibinfo
  {volume} {101}},\ \bibinfo {pages} {020401} (\bibinfo {year} {2008})},\
  \Eprint {http://arxiv.org/abs/0710.5549} {arXiv:0710.5549} \BibitemShut
  {NoStop}%
\bibitem [{\citenamefont {Tollaksen}(2007)}]{tollaksen}%
  \BibitemOpen
  \bibfield  {author} {\bibinfo {author} {\bibfnamefont {J.}~\bibnamefont
  {Tollaksen}},\ }\href {\doibase 10.1088/1751-8113/40/30/025} {\bibfield
  {journal} {\bibinfo  {journal} {J. Phys. A: Math. Theor.}\ }\textbf {\bibinfo
  {volume} {40}},\ \bibinfo {pages} {9033} (\bibinfo {year} {2007})},\ \Eprint
  {http://arxiv.org/abs/quant-ph/0602226} {arXiv:quant-ph/0602226} \BibitemShut
  {NoStop}%
\bibitem [{\citenamefont {Jozsa}(2007)}]{jozsa}%
  \BibitemOpen
  \bibfield  {author} {\bibinfo {author} {\bibfnamefont {R.}~\bibnamefont
  {Jozsa}},\ }\href {\doibase 10.1103/PhysRevA.76.044103} {\bibfield  {journal}
  {\bibinfo  {journal} {Phys. Rev. A}\ }\textbf {\bibinfo {volume} {76}},\
  \bibinfo {pages} {044103} (\bibinfo {year} {2007})},\ \Eprint
  {http://arxiv.org/abs/0706.4207} {arXiv:0706.4207} \BibitemShut {NoStop}%
\bibitem [{\citenamefont {Bartlett}\ \emph {et~al.}(2012)\citenamefont
  {Bartlett}, \citenamefont {Rudolph},\ and\ \citenamefont {Spekkens}}]{erl}%
  \BibitemOpen
  \bibfield  {author} {\bibinfo {author} {\bibfnamefont {S.~D.}\ \bibnamefont
  {Bartlett}}, \bibinfo {author} {\bibfnamefont {T.}~\bibnamefont {Rudolph}}, \
  and\ \bibinfo {author} {\bibfnamefont {R.~W.}\ \bibnamefont {Spekkens}},\
  }\href {\doibase 10.1103/PhysRevA.86.012103} {\bibfield  {journal} {\bibinfo
  {journal} {Phys. Rev. A}\ }\textbf {\bibinfo {volume} {86}},\ \bibinfo
  {pages} {012103} (\bibinfo {year} {2012})},\ \Eprint
  {http://arxiv.org/abs/1111.5057} {arXiv:1111.5057} \BibitemShut {NoStop}%
\bibitem [{\citenamefont {Kochen}\ and\ \citenamefont {Specker}(1968)}]{ks}%
  \BibitemOpen
  \bibfield  {author} {\bibinfo {author} {\bibfnamefont {S.}~\bibnamefont
  {Kochen}}\ and\ \bibinfo {author} {\bibfnamefont {E.}~\bibnamefont
  {Specker}},\ }\href {\doibase 10.1512/iumj.1968.17.17004} {\bibfield
  {journal} {\bibinfo  {journal} {Indiana Univ. Math. J.}\ }\textbf {\bibinfo
  {volume} {17}},\ \bibinfo {pages} {59} (\bibinfo {year} {1968})}\BibitemShut
  {NoStop}%
\bibitem [{\citenamefont {Spekkens}(2014)}]{robrant}%
  \BibitemOpen
  \bibfield  {author} {\bibinfo {author} {\bibfnamefont {R.~W.}\ \bibnamefont
  {Spekkens}},\ }\href {\doibase 10.1007/s10701-014-9833-x} {\bibfield
  {journal} {\bibinfo  {journal} {Found. Phys.}\ }\textbf {\bibinfo {volume}
  {44}},\ \bibinfo {pages} {1125} (\bibinfo {year} {2014})},\ \Eprint
  {http://arxiv.org/abs/1312.3667} {arXiv:1312.3667} \BibitemShut {NoStop}%
\bibitem [{\citenamefont {Nielsen}\ and\ \citenamefont {Chuang}(2000)}]{nc}%
  \BibitemOpen
  \bibfield  {author} {\bibinfo {author} {\bibfnamefont {M.~A.}\ \bibnamefont
  {Nielsen}}\ and\ \bibinfo {author} {\bibfnamefont {I.~L.}\ \bibnamefont
  {Chuang}},\ }\href@noop {} {\emph {\bibinfo {title} {Quantum Computation and
  Quantum Information}}}\ (\bibinfo  {publisher} {Cambridge University Press},\
  \bibinfo {year} {2000})\BibitemShut {NoStop}%
\bibitem [{Note1()}]{Note1}%
  \BibitemOpen
  \bibinfo {note} {Notice that although $p_-$ is a combination of operationally
  defined quantities, it is not exactly the probability of getting a negative
  $x$ under the pre- and post-selection. To obtain this, instead of dividing by
  $p_\phi $ one would have to divide by $\protect \Braket {\psi |S|\psi }=
  (1-p_d)p_\phi + p_d\protect \Braket {\psi |E_d|\psi }$, making the analysis
  slightly more complicated (but still tractable).}\BibitemShut {Stop}%
\bibitem [{\citenamefont {Garretson}\ \emph {et~al.}(2004)\citenamefont
  {Garretson}, \citenamefont {Wiseman}, \citenamefont {Pope},\ and\
  \citenamefont {Pegg}}]{wiseweak}%
  \BibitemOpen
  \bibfield  {author} {\bibinfo {author} {\bibfnamefont {J.~L.}\ \bibnamefont
  {Garretson}}, \bibinfo {author} {\bibfnamefont {H.~M.}\ \bibnamefont
  {Wiseman}}, \bibinfo {author} {\bibfnamefont {D.~T.}\ \bibnamefont {Pope}}, \
  and\ \bibinfo {author} {\bibfnamefont {D.~T.}\ \bibnamefont {Pegg}},\ }\href
  {\doibase 10.1088/1464-4266/6/6/008} {\bibfield  {journal} {\bibinfo
  {journal} {J. Opt. B: Quantum Semiclass. Opt.}\ }\textbf {\bibinfo {volume}
  {6}},\ \bibinfo {pages} {S506} (\bibinfo {year} {2004})},\ \Eprint
  {http://arxiv.org/abs/quant-ph/0310081} {arXiv:quant-ph/0310081} \BibitemShut
  {NoStop}%
\bibitem [{Note2()}]{Note2}%
  \BibitemOpen
  \bibinfo {note} {This is because the analysis presented here, like any proof
  of contextuality, is for an ontological model that produces individual
  measurement results with the correct probabilities. This requirement
  immediately rules out a straightforward field ontology that, whilst perhaps
  offering interesting explanations of the weak value \cite
  {classical2,field1,field2}, only produces intensities. To exclude
  non-contextual explanation, an experiment based on fields would have to
  justify this requirement by working at the level of single field quanta.
  Compare with the classic ``double-slit experiment'': whilst an interference
  pattern in intensities has a simple classical explanation in terms of fields,
  the same interference pattern in what are unambiguously probabilities defies
  classical intuitions}\BibitemShut {NoStop}%
\bibitem [{\citenamefont {Spekkens}\ and\ \citenamefont {Kunjwal}()}]{ravi}%
  \BibitemOpen
  \bibfield  {author} {\bibinfo {author} {\bibfnamefont {R.~W.}\ \bibnamefont
  {Spekkens}}\ and\ \bibinfo {author} {\bibfnamefont {R.}~\bibnamefont
  {Kunjwal}},\ }\href@noop {} {\enquote {\bibinfo {title} {Operational
  inequality for noncontextuality in the {S}pecker scenario},}\ }\bibinfo
  {note} {(in preparation), see also
  \href{http://pirsa.org/14010102/}{PIRSA:14010102}}\BibitemShut {NoStop}%
\bibitem [{\citenamefont {Berry}\ and\ \citenamefont {Popescu}(2006)}]{field1}%
  \BibitemOpen
  \bibfield  {author} {\bibinfo {author} {\bibfnamefont {M.~V.}\ \bibnamefont
  {Berry}}\ and\ \bibinfo {author} {\bibfnamefont {S.}~\bibnamefont
  {Popescu}},\ }\href {http://stacks.iop.org/0305-4470/39/i=22/a=011}
  {\bibfield  {journal} {\bibinfo  {journal} {J. Phys. A: Math. Gen.}\ }\textbf
  {\bibinfo {volume} {39}},\ \bibinfo {pages} {6965} (\bibinfo {year}
  {2006})}\BibitemShut {NoStop}%
\bibitem [{\citenamefont {Dressel}\ \emph
  {et~al.}(2014{\natexlab{b}})\citenamefont {Dressel}, \citenamefont {Bliokh},\
  and\ \citenamefont {Nori}}]{field2}%
  \BibitemOpen
  \bibfield  {author} {\bibinfo {author} {\bibfnamefont {J.}~\bibnamefont
  {Dressel}}, \bibinfo {author} {\bibfnamefont {K.~Y.}\ \bibnamefont {Bliokh}},
  \ and\ \bibinfo {author} {\bibfnamefont {F.}~\bibnamefont {Nori}},\ }\href
  {\doibase 10.1103/PhysRevLett.112.110407} {\bibfield  {journal} {\bibinfo
  {journal} {Phys. Rev. Lett.}\ }\textbf {\bibinfo {volume} {112}},\ \bibinfo
  {pages} {110407} (\bibinfo {year} {2014}{\natexlab{b}})},\ \Eprint
  {http://arxiv.org/abs/1308.4831} {arXiv:1308.4831} \BibitemShut {NoStop}%
\end{thebibliography}%
\end{document}